\documentclass[conference]{IEEEtran}
\IEEEoverridecommandlockouts

\usepackage{cite}
\usepackage{amsmath,amssymb,amsfonts}
\usepackage{algorithmic}
\usepackage{graphicx}
\usepackage{textcomp}
\usepackage{newfloat}
\usepackage{listings}
\usepackage{booktabs}
\usepackage{multirow}
\usepackage{xcolor}
\usepackage{soul}
\usepackage{makecell}
\usepackage{amsmath}
\usepackage{amsfonts}
\usepackage{times}
\usepackage{helvet}
\usepackage{courier}
\usepackage{caption}
\def\BibTeX{{\rm B\kern-.05em{\sc i\kern-.025em b}\kern-.08em
    T\kern-.1667em\lower.7ex\hbox{E}\kern-.125emX}}
\begin{document}

\title{IKMo: Image-Keyframed Motion Generation with Trajectory-Pose Conditioned Motion Diffusion Model}

\author{\IEEEauthorblockN{Yang Zhao$^{1}$, Yan Zhang$^{1}$, Zixuan Guo$^{1}$, Hangyu Zhou$^{1,2}$, Haotian Mao$^{1}$, Hui Wang$^{1}$, Xubo Yang$^{1\dagger}$}
\IEEEauthorblockA{$^{1}$\textit{Shanghai Jiao Tong University}, China; $^{2}$\textit{Zhejiang University of Science and Technology}, China\\
runder1103@sjtu.edu.cn, yan-zh@sjtu.edu.cn, zixuanguo@sjtu.edu.cn\\
zhy@zust.edu.cn, maohaotian@sjtu.edu.cn, wanghehv@sjtu.edu.cn, yangxubo@sjtu.edu.cn}
\thanks{$^{\dagger}$Corresponding author: Xubo Yang. Supported by the Key Technology Research and Development Program of Shanghai (25JC3202303).}}

\IEEEaftertitletext{\vspace{-24pt}}
\maketitle

\begin{abstract}
While text-to-motion generation is mainstream, integrating explicit spatial constraints has become a standard solution for precise animation. Existing state-of-the-art models are tailored for professional workflows, effectively processing high-quality, pre-aligned 3D inputs via unified conditioning mechanisms. However, this strictly aligned setting restricts accessibility for general users who rely on unaligned, in-the-wild 2D reference images. This accessible scenario introduces a fundamental conflict: heterogeneous geometric constraints, where user-defined global trajectories clash with local image-derived poses. To resolve this coordinate mismatch, we propose TPMDM, a diffusion model featuring a decoupled parallel control strategy that explicitly isolates and aligns these divergent signals. Furthermore, we introduce IKMo, a holistic framework that synergizes TPMDM with a VLM-powered multi-agent system. By bridging the semantic-geometric gap, IKMo translates user-friendly inputs into explicit motion constraints to drive the diffusion backbone. Experiments on HumanML3D and KIT-ML demonstrate that our method establishes a new state-of-the-art in the task of generating motions under heterogeneous trajectory and keyframe pose constraints.
\end{abstract}


\section{Introduction}
\label{sec:intro}

Human motion generation is a core task in computer animation, enabling virtual characters in applications such as gaming, film, and virtual reality. Recently, text-to-motion methods have gained popularity for generating natural motions from language descriptions.

To enable more explicit and precise control, integrating explicit constraints—specifically global trajectories and 3D pose keyframes—has become a standard solution. State-of-the-art methods like OmniControl~\cite{xie2023omnicontrol} and CondMDI~\cite{cohan2024flexible} have established robust paradigms for this task. These models effectively employ a unified mechanism to process these multi-modal signals, assuming the availability of high-quality, pre-aligned 3D constraints typically found in professional animation workflows.

\begin{figure}
  \includegraphics[width=1\linewidth]{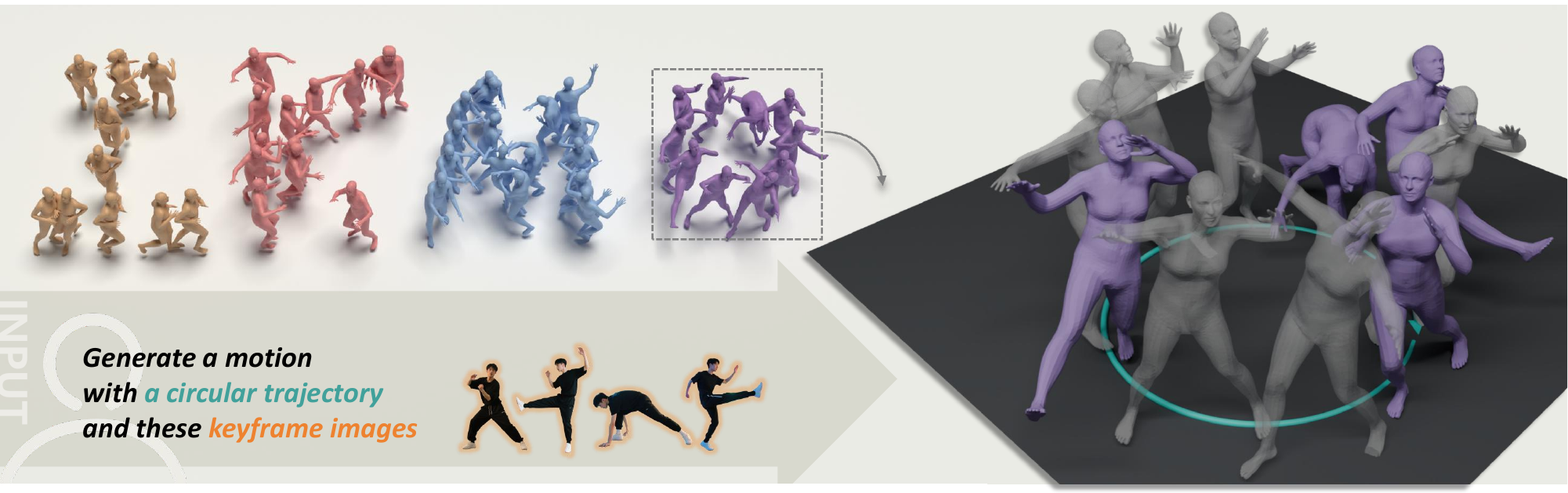}
  \caption{Our diffusion model jointly controls trajectories and keyframe poses, while IKMo adds a multi-agent MLLM interface for image-guided generation.}
  \label{fig:myteaser}
\end{figure}

However, a distinct challenge arises when expanding the scope to general users who lack access to curated 3D assets. A more accessible paradigm is to extract structural cues directly from in-the-wild 2D reference images. Unlike the strictly aligned inputs in professional pipelines, this accessible setting introduces a fundamental geometric conflict: heterogeneous geometric constraints. Specifically, user-defined trajectories represent absolute global paths, whereas poses derived from unaligned images provide only relative local structures. Existing unified mechanisms, which treat these divergent signals indiscriminately as homogeneous constraints, struggle to reconcile such coordinate discrepancies. Without an explicit design to handle this heterogeneity, the optimization process faces conflicting gradients, often leading to generation artifacts.

We propose TPMDM, a motion diffusion model that incorporates both global trajectory and local keyframe pose constraints for controllable human motion generation. To effectively handle these spatially heterogeneous signals, we decouple the two modalities and process them via parallel control pathways. This design mitigates gradient conflicts, enabling robust and flexible motion control. Experiments on HumanML3D and KIT-ML demonstrate that our model outperforms state-of-the-art methods across all metrics under both constraint types.

While TPMDM effectively resolves spatial conflicts, a cognitive barrier remains: bridging the semantic-geometric gap between abstract user intents and explicit 3D constraints. For general users, mentally mapping high-level descriptions to precise 3D trajectories and pose sequences is cognitively demanding. To overcome this interaction-level challenge, we introduce IKMo, a holistic framework that synergizes a VLM-powered multi-agent system with our decoupled diffusion backbone. Within this framework, the multi-agent module acts as an intelligent interpreter, effectively grounding high-level semantics into precise motion synthesis signals. These parsed signals subsequently drive TPMDM, enabling the intuitive generation of highly controllable motion from abstract multimodal inputs.

The main contributions are summarized as follows:

\begin{itemize}
    \item We propose \textbf{TPMDM}, a motion diffusion model that employs a decoupled parallel control strategy to explicitly resolve the spatial coordinate conflict between global trajectories and local image-based poses.
    \item We introduce \textbf{IKMo}, a VLM-powered framework designed to bridge the semantic-geometric gap, enabling intuitive motion generation by interpreting abstract multimodal user intents into explicit geometric constraints.
    \item Our method sets a new state-of-the-art on the HumanML3D and KIT-ML datasets for motion generation tasks conditioned on heterogeneous trajectory-keyframe poses.
\end{itemize}

\section{Related Work}

Human motion generation aims to synthesize natural motion sequences under various control conditions, which can be categorized by modality.
Text-to-motion~\cite{ tevethuman, petrovich2023tmr, jiang2023motiongpt, guo2024momask, zhang2023remodiffuse, zhang2024motiondiffuse} uses language descriptions as input.
However, such linguistic descriptions are inherently ambiguous and impose relatively weak constraints on the motion.

To enable fine-grained manipulation, explicit spatial constraints have been integrated into the generation process.
Trajectory-guided methods~\cite{kaufmann2020convolutional, karunratanakul2023guided, rempe2023trace, shafir2024human, wan2024tlcontrol} use predefined motion paths, while keyframe-guided methods~\cite{xie2023omnicontrol, cohan2024flexible, hwang2025motionsynthesissparseflexible} impose spatial or pose constraints at specific frames.
Recent works have also explored visual modalities, such as sketches~\cite{wu2024doodle}, to provide more intuitive structural guidance.

Crucially, OmniControl~\cite{xie2023omnicontrol} advances this field by enabling versatile control through arbitrary joint constraints, allowing users to spatially pin any joint at any frame. Similarly, CondMDI~\cite{cohan2024flexible} utilizes sparse keyframes for motion interpolation.
Despite these capabilities, manually specifying precise 3D coordinates for trajectories or arbitrary joints remains labor-intensive and unintuitive for non-experts. A logical solution is to leverage 2D reference images, extracting 3D structural cues via off-the-shelf estimators to guide generation.
However, integrating this accessible modality poses a challenge for existing spatial control methods like OmniControl~\cite{xie2023omnicontrol} and CondMDI~\cite{cohan2024flexible}. Designed for high-precision controllable generation, these models typically employ a unified conditioning strategy, treating all spatial signals---whether global trajectories or poses---as homogeneous absolute constraints. This approach is highly effective when inputs are pre-aligned, as it optimizes generation by enforcing strict alignment between the output and the control signals in a shared coordinate space.

However, this assumption holds only within professional workflows. When fusing global trajectories with unaligned, image-derived 3D poses, this unified mechanism induces severe constraint conflicts. The model struggles to reconcile the mismatch between the global navigation requirements and the local relative pose structures, often leading to generation artifacts. To address this, we propose a decoupled strategy that processes these heterogeneous constraints explicitly and independently.

\begin{figure*}[ht]
    \centering
    \includegraphics[width=1\linewidth]{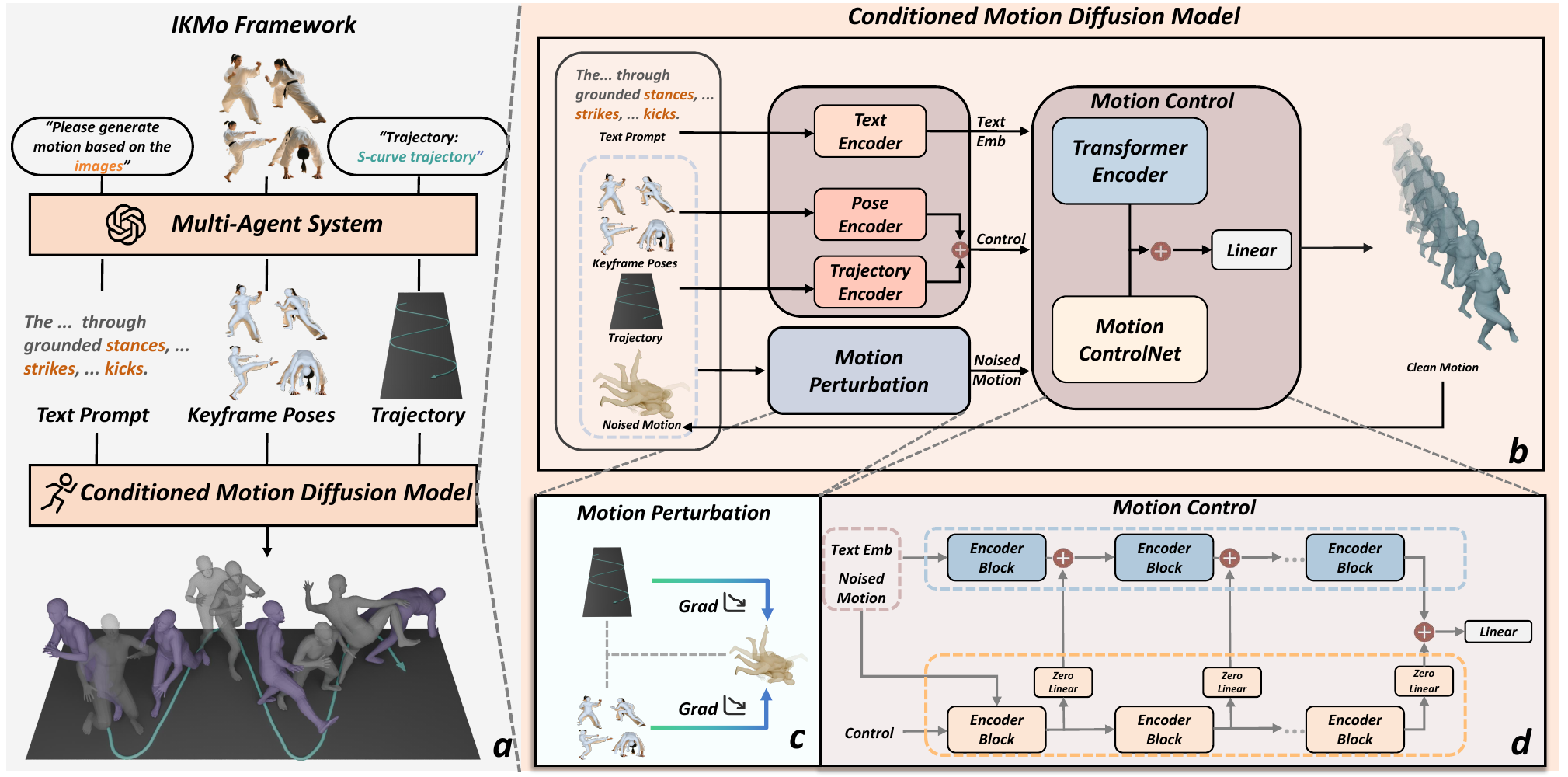}
    \caption{
    \textbf{(\textit{a}) IKMo pipeline.} A multi-agent MLLM extracts keyframes and trajectories from image–text input.
    \textbf{(\textit{b}) Diffusion model.} Generates motion from noise under these constraints.
    \textbf{(\textit{c}) Perturbation.} Iteratively refines the result.
    \textbf{(\textit{d}) Control.} Trajectory and keyframe features are fused via ControlNet. }

    \label{fig:pipeline}
\end{figure*}

\section{METHOD}

Existing controllable motion diffusion models implicitly assume that all spatial constraints can be represented and optimized in a unified global coordinate system.

In this work, we argue that this assumption no longer holds in in-the-wild scenarios, as the global trajectories and the poses derived from reference images are inherently unaligned.
We propose \textbf{TPMDM} (Figure~\ref{fig:pipeline}\textit{b–d}), a decoupled diffusion model that processes them separately for improved control and alignment.
Building on this, \textbf{IKMo} (Figure~\ref{fig:pipeline}\textit{a}) employs a multi-agent system to convert image–text prompts into motion instructions, which TPMDM synthesizes into coordinated motion.
This section details trajectory and pose guidance, their integration into the denoising network, and image-based keyframe generation.

\subsection{Motion Generation With Trajectory Control}

Inspired by OmniControl~\cite{xie2023omnicontrol}, we adopt a two-stage trajectory guidance framework that effectively incorporates sparse trajectory constraints and serves as a strong foundation for our method.

In the first stage, \textit{Trajectory Perturbation}, an analytical function $L_{\text{traj}}(x, c_{\text{t}})$ computes the L2 distance between joint positions and trajectory constraints, guiding generation via gradient-based perturbation. We constrain only the pelvis joint to define the global trajectory:

\begin{equation}
L_{\text{traj}}(x, c_{\text{t}}) = \frac{\sum_n \sigma_{n;\text{t}} \left\lVert c_{n;\text{t}} - x_{\text{n}, root}^g \right\rVert_2}{\sum_n \sigma_{n;\text{t}}}, \quad x^g = R(x),
\end{equation}
where $\sigma_{n;\text{t}}$ is a binary indicator that specifies whether the trajectory control signal $c_{n;\text{t}}$ contains a valid value at frame $n$ for the root joint, and $R(\cdot)$ transforms local joint coordinates into global absolute positions.

However, since this guidance strategy is applied to only a single joint, it fails to effectively propagate spatial constraints to the rest of the body through backpropagation, often resulting in unrealistic or physically implausible full-body configurations. To mitigate this limitation, a second stage is introduced. This stage consists of a trajectory encoder and a trainable copy of the Transformer encoder—composed of stacked encoder blocks—originally used in the motion diffusion model. By conditioning on the spatial signals, it facilitates the generative model in learning globally coherent full-body motion patterns.

\subsection{Motion Generation With Keyframe Pose Control}

Pose guidance ensures consistency with keyframe poses, implemented via a two-stage strategy in our framework.

In the \textit{Pose Perturbation Module}, we introduce a pose analysis function to quantify the alignment between generated and target poses.
\begin{equation}
L_{\text{pose}}(x, c_{\text{p}}) = \frac{\sum_n \sigma_{n;\text{p}} \left\lVert Align(x_{\text{n}}^g, c_{n;\text{p}}, c_{n;\text{t}}) - x_{\text{n}}^g \right\rVert_2}{\sum_n \sigma_{n;\text{p}}},
\end{equation}

\noindent where $\sigma_{n;\text{p}}$ is a binary indicator that specifies whether the pose control signal $c_{n;\text{p}}$ provides a valid constraint at frame $n$. The function $L_{\text{pose}}(\cdot)$ evaluates the L2 distance between the motion pose and the aligned constraint pose.

To enhance sensitivity to relative pose structure, we apply spatial alignment to constraint poses. For frames with trajectory constraints, the pose is translated to match the root joint with the constraint position; otherwise, it is aligned to the root projection of the corresponding motion frame. This mitigates global displacement and enables perturbation to focus on local pose differences.

The alignment function is denoted as:

\begin{equation}
\text{Align}(x^g_n, c_{n;\text{p}}, c_{n;\text{t}})
= c_{n;\text{p}} + \Delta_{\text{A}},
\end{equation}

\begin{equation}
\Delta_{\text{A}} =
\begin{cases}
\text{Pos}(c_{n;\text{t}, \text{root}}) - \text{Pos}(c_{n;\text{p}, \text{root}}),
&  \sigma_{n;\text{t}} = 1 \\
\left(
\begin{aligned}
& \text{Proj}_x(x^g_{n, \text{root}}) - \text{Proj}_x(c_{n;\text{p}, \text{root}}), \\
& 0, \\
& \text{Proj}_z(x^g_{n, \text{root}}) - \text{Proj}_z(c_{n;\text{p}, \text{root}})
\end{aligned}
\right), & \sigma_{n;\text{t}} = 0
\end{cases}
\end{equation}

\noindent where $Pos(\cdot)$ represents the pelvis joint coordinates of the frame's motion, $Proj_x(\cdot)$ represents the projection of the pelvis coordinates onto the x-axis, and $Proj_z(\cdot)$ represents the projection of the pelvis coordinates onto the z-axis.

To further improve motion style accuracy and temporal consistency, we introduce a second-stage pose guidance module comprising a pose encoder and a Motion ControlNet. Conditioned on keyframe poses, it runs in parallel with the diffusion model to provide continuous guidance. This enhances frame-wise fidelity and enforces semantic and structural coherence across the sequence.

\subsection{Synergistic Guidance via Trajectory and Keyframe Poses}
\label{sec:both}
We randomly sample trajectories and pose sequences along the temporal dimension. The trajectory is represented by a sequence of root joint coordinates, while the pose corresponds to a sequence of full-body spatial joint positions. To mitigate inconsistencies in global rotation between image-based 3D poses and real-world motions, we apply small-scale global random rotations to the sampled poses before training.

In the first stage, \textit{Motion Perturbation}, we iteratively apply gradient descent at each diffusion timestep to update the noised motion $x$, based on a combined loss from the pose and trajectory analysis functions. The guidance process is shown in Figure~\ref{fig:pipeline}(\textit{c}), where "Noised motion" denotes $x$. The perturbation is defined as:
\begin{equation}
x = x - \tau \nabla_{x} \left[ \lambda \cdot L_{\text{traj}}(x, c_{\text{t}}) + (1 - \lambda) \cdot L_{\text{pose}}(x, c_{\text{p}}) \right],
\end{equation}
where $\tau$ denotes the guidance strength, and the weighting factor $\lambda$ is dynamically computed based on the relative loss magnitudes:
\begin{equation}
\lambda = \frac{L_{\text{traj}}(x, c_{\text{t}})}{L_{\text{traj}}(x, c_{\text{t}}) + L_{\text{pose}}(x, c_{\text{p}})},
\end{equation}
which reflects the relative contributions of trajectory and pose errors. This joint guidance allows the model to balance structural coherence and spatial alignment adaptively.

In the second stage, we employ a \textit{Motion ControlNet}, a trainable copy of the Transformer encoder used in the motion diffusion model. It incorporates trajectory and keyframe pose constraints, which are independently processed through dedicated trajectory and pose encoders. The encoded features are fused and injected into the ControlNet, allowing the integration of both spatial and semantic information into the denoising backbone. Crucially, ControlNet itself does not resolve the geometric conflict between trajectory and pose; it only propagates constraint features once such conflicts are mitigated by decoupled perturbation. The guidance process is illustrated in Figure~\ref{fig:pipeline}(\textit{b,d}).

\subsection{Image Keyframe Controlled Motion Generation}

To enable motion generation from visual and textual inputs, we propose a multi-agent system with three specialized agents: an \textit{Interaction Agent}, a \textit{Motion Design Agent}, and a \textit{Trajectory Planning Agent}. This module serves as an interface that translates user-friendly image and text inputs into explicit motion constraints. Unlike single-agent LLMs, this collaborative design improves multimodal parsing and structured planning. The Interaction Agent extracts user intent, the Motion Design Agent recovers 3D keyframe poses from images via TokenHMR~\cite{dwivedi2024tokenhmr} and maps them to a canonical SMPL~\cite{loper2015smpl} joint format aligned with HumanML3D~\cite{guo2022generating}, and the Trajectory Planning Agent generates parameterized 3D paths. Together, these agents produce motion specifications—structured prompts, aligned poses, and trajectories—that condition our diffusion model for flexible synthesis.

\begin{table*}[t]
\caption{\textbf{Quantitative results on the HumanML3D test set.} \textbf{Bold} indicates the best result, \underline{underline} the second-best. $\uparrow$ means higher is better, $\downarrow$ lower is better, and $\rightarrow$ closer to real data is preferred.}
\centering
\scriptsize
\setlength{\tabcolsep}{3pt}
\resizebox{\linewidth}{!}{
\begin{tabular}{llccccccc}
\toprule
\textbf{Method} & \textbf{Condition} & \textbf{FID} $\downarrow$ &
\makecell{\textbf{R-precision} \\ (Top-3) $\uparrow$} &
\textbf{Diversity} $\rightarrow$ &
\makecell{\textbf{Foot skating} \\ ratio $\downarrow$} &
\makecell{\textbf{Traj.err.} \\ (50 cm) $\downarrow$} &
\makecell{\textbf{Loc.err.} \\ (50 cm) $\downarrow$} &
\textbf{Avg.err. } $\downarrow$ \\
\midrule
Real & - & 0.002 & 0.797 & 9.503 & 0.000 & 0.000 & 0.000 & 0.000 \\
\midrule
MDM & Pelvis & 0.698 & 0.602 & 9.197 & 0.1019 & 0.4022 & 0.3076 & 0.5959 \\
PriorMDM & Pelvis & 0.475 & 0.583 & 9.156 & 0.0897 & 0.3457 & 0.2132 & 0.4417 \\
GMD & Pelvis & 0.576 & 0.665 & 9.206 & 0.1009 & 0.0931 & 0.0321 & 0.1439 \\
\midrule
OmniControl & Pelvis+Keyframes & 1.420 & 0.652 & 9.268 & 0.1039 & 0.4457 & 0.0670 & 0.1371 \\
CondMDI & Pelvis+Keyframes & 0.266 & \textbf{0.678} & 9.198 & 0.0900 & 0.4671 & 0.2093 & 0.1661 \\
\textbf{Ours(undecoupled)} & Pelvis+Keyframes & 0.262 & 0.671 & \underline{9.677} & 0.0821 & 0.0510 & 0.0153 & 0.0500 \\
\textbf{Ours(w/ rotations)} & Pelvis+Keyframes & \underline{0.239} & \textbf{0.678} & 9.686 & \textbf{0.0543} & \underline{0.0246} & \underline{0.0076}& \underline{0.0250}\\
\textbf{Ours(w/o rotations)} & Pelvis+Keyframes & \textbf{0.177} & \underline{0.672} & \textbf{9.616} & \underline{0.0609} & \textbf{0.0176} & \textbf{0.0056}& \textbf{0.0212}\\
\bottomrule
\end{tabular}
}

\label{tab:comparison_humanml3d}
\end{table*}

\begin{table*}[t]
\caption{\textbf{Quantitative results on the KIT-ML test set.} }
\centering
\small
\begin{tabular}{llcccccc}
\toprule
\textbf{Method} & \textbf{Condition} & \textbf{FID} $\downarrow$ &
\makecell{\textbf{R-precision} \\ (Top-3) $\uparrow$} &
\textbf{Diversity} $\rightarrow$ &
\makecell{\textbf{Traj.err.} \\ (50 cm) $\downarrow$} &
\makecell{\textbf{Loc.err.} \\ (50 cm) $\downarrow$} &
\textbf{Avg.err.} $\downarrow$ \\
\midrule
Real & - & 0.031 & 0.779 & 11.08 & 0.000 & 0.000 & 0.000 \\
\midrule
PriorMDM & Pelvis & 0.851 & 0.397 & 10.518 & 0.3310 & 0.1400 & 0.2305 \\
GMD & Pelvis & 1.565 & 0.382 & 9.664 & 0.5443 & 0.3003 & 0.4070 \\

\midrule
OmniControl & Pelvis+Keyframes & \underline{0.690} & \underline{0.405} & \underline{10.597} & \underline{0.1226} & \underline{0.0123} & \underline{0.0673} \\
\textbf{Ours} & Pelvis+Keyframes & \textbf{0.531} & \textbf{0.413} & \textbf{10.748} & \textbf{0.0341} & \textbf{0.0084} & \textbf{0.0333} \\
\bottomrule

\end{tabular}

\label{tab:comparison_kit}
\end{table*}

\begin{table}[htbp]
\caption{\textbf{Pose Dist Evaluation on the HumanML3D test set.} }
\centering
\small
\begin{tabular}{lcc}
\toprule
\textbf{Method} & \textbf{Condition} & \textbf{Pose Dist} $\downarrow$ \\
\midrule
Real & - & 0.000 \\
\midrule
OmniControl & Pelvis+Keyframes & 0.0373 \\
\textbf{Ours(undecoupled)} & Pelvis+Keyframes & 0.0249 \\
\textbf{Ours} & Pelvis+Keyframes & \textbf{0.0210} \\
\bottomrule
\end{tabular}

\label{tab:posedist_humanml3d}
\end{table}

\section{RESULTS}

\begin{figure}[ht]
    \centering
    \includegraphics[width=1\linewidth]{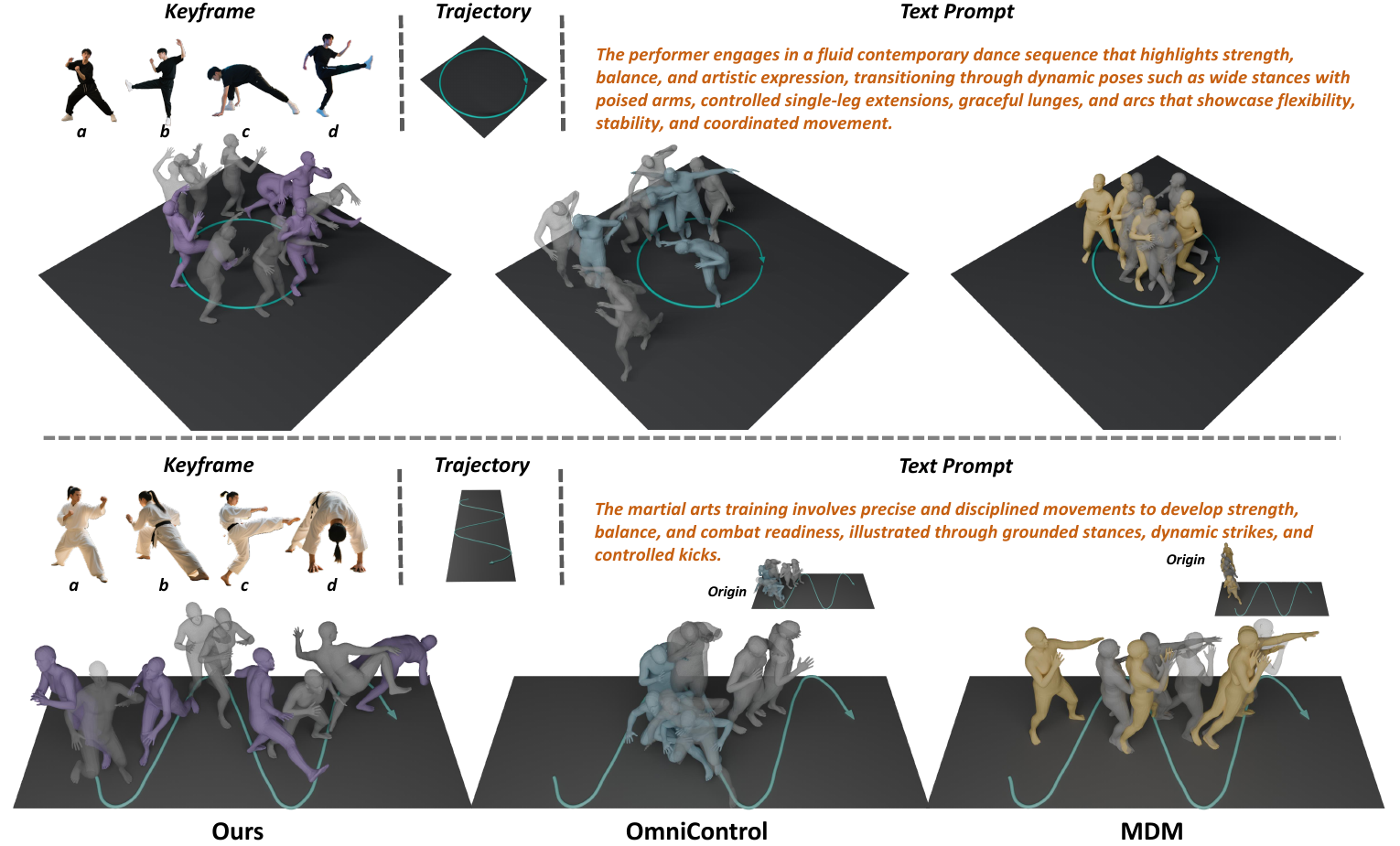}
    \caption{\textbf{Qualitative Results.} Colored frames are keyframes and gray frames are non-keyframes, with transparency indicating temporal order. Green curves denote reference trajectories, and \textit{Origin} shows source motions. Some results are rigidly aligned for clearer visualization.}
    \label{fig:result}
\end{figure}

\subsection{Datasets and Evaluation Metrics}

We evaluate our model on the HumanML3D~\cite{guo2022generating} dataset, which contains 14,646 text-annotated motion sequences sourced from AMASS~\cite{mahmood2019amass} and HumanAct12~\cite{guo2020action2motion}. Additionally, we evaluate on the KIT-ML~\cite{plappert2016kit} dataset comprising 3,911 sequences.

We follow the evaluation protocol proposed in~\cite{guo2022generating}, using FID to measure motion realism, R-Precision for text-motion alignment, and Diversity for intra-sample variation. To assess control accuracy, we report Foot skating ratio, Trajectory Error, Location Error, and Average Error at keyframes~\cite{karunratanakul2023guided}. To further quantify the accuracy of pose control, we introduce a new metric, \textit{Pose Dist}, which evaluates the average Euclidean distance between pelvis-centered generated and reference poses at controlled frames. In all quantitative experiments, we randomly sampled the same number of trajectory points and keyframes, while for qualitative comparisons, we ensured identical trajectories, keyframe poses, and temporal alignment. Full implementation details are provided in the supplementary material.

\subsection{Quantitative Evaluation}

We compare our method against both mainstream and SOTA approaches on the HumanML3D~\cite{guo2022generating} and KIT-ML~\cite{plappert2016kit} datasets. Table~\ref{tab:comparison_humanml3d} presents the performance comparison on HumanML3D, while Table~\ref{tab:comparison_kit} shows the results on KIT-ML.

As shown in Table~\ref{tab:comparison_humanml3d}, prior methods such as MDM~\cite{tevethuman}, PriorMDM~\cite{shafir2024human}, and GMD~\cite{karunratanakul2023guided} provide only trajectory-level control, while OmniControl~\cite{xie2023omnicontrol} extends this to both trajectory and pose-level control. CondMDI~\cite{cohan2024flexible} represents the SOTA in keyframe-guided motion generation, supporting both trajectory and pose conditions.

We compare our method against these approaches and additionally analyze the effect of rotation-augmented training. Since our model is designed to handle arbitrary global rotations, the rotation-trained version (Ours (w/ rotations)) is adopted in all subsequent experiments. Results on both HumanML3D and KIT-ML (Table~\ref{tab:comparison_kit}) demonstrate that our approach consistently outperforms prior methods across all metrics.

Since OmniControl adopts a similar pose representation with us, it is compatible with our proposed Pose Dist metric. We evaluate this metric on the HumanML3D dataset, as presented in Table~\ref{tab:posedist_humanml3d}. The results demonstrate that our method achieves better keyframe pose similarity.

\subsection{Qualitative Results}

Figure~\ref{fig:result} qualitatively compares our method with OmniControl~\cite{xie2023omnicontrol} and MDM~\cite{tevethuman}.
All methods use the same keyframes and trajectories for fairness; specifically, for MDM, we append additional text prompts to constrain the trajectory and pose. Our model accurately follows both keyframe poses and trajectories, preserving motion style and spatial consistency. In contrast, OmniControl suffers from pose mismatches and trajectory deviation due to its joint-level formulation, which rigidly interprets unaligned pose keyframes as absolute spatial constraints, leading to severe coordinate conflicts with the global trajectory. Meanwhile, MDM largely ignores the provided trajectory and keyframe prompts. For clarity, results of baselines are translated and rotated, with original videos provided in the supplementary material.

\begin{figure}[htbp]
    \centering
    \includegraphics[width=1\linewidth]{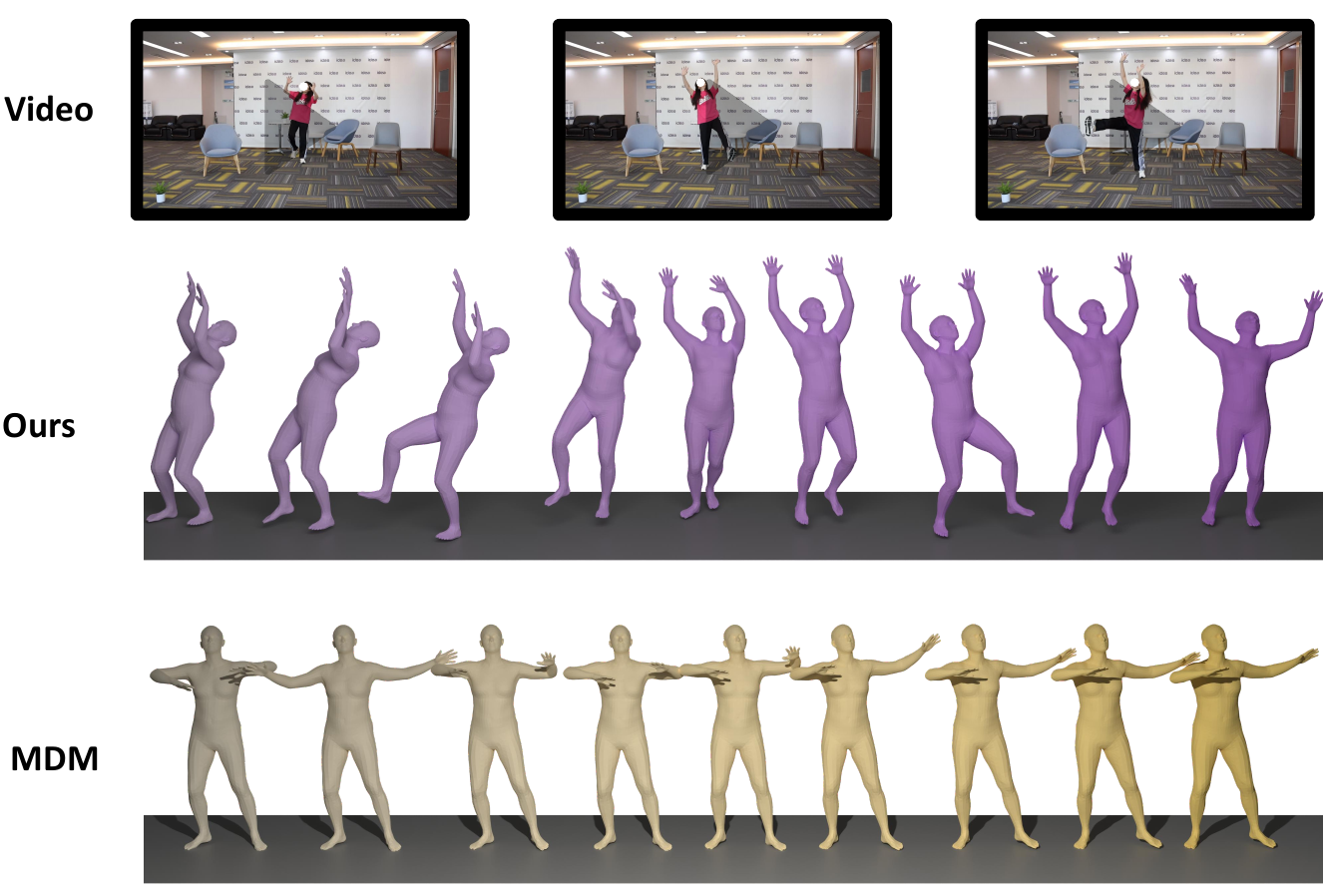}
    \caption{\textbf{Qualitative results using video/text inputs.} Both methods are given the same textual prompt.}
    \label{fig:video_result}
\end{figure}

Moreover, for video-based motions, IKMo effectively captures motion style using keyframes extracted from video. As shown in Figure~\ref{fig:video_result}, we compare with MDM using video and text from the MoVid dataset~\cite{chen2024motionllm}. IKMo takes both text and video keyframes as input, while MDM uses text only. Our method better replicates the reference style, whereas MDM yields generic arm-swinging motions. Unlike pose estimation methods, IKMo supports flexible keyframe sparsity, enabling user-adjustable similarity and diverse motion generation.

\subsection{Ablation Study}

We conduct ablation studies on the HumanML3D~\cite{guo2022generating} dataset to validate the effectiveness of the proposed modules in our motion diffusion model. We summarize several key findings below.

\begin{table}[htbp]
\caption{\textbf{Ablation study on the control modules.} }
\centering
\small
\setlength{\tabcolsep}{3.5pt}
\begin{tabular*}{\linewidth}{@{\extracolsep{\fill}} llccc}
\toprule
\textbf{Method} &
\makecell{\textbf{Traj.err.} \\ (50 cm) $\downarrow$} &
\makecell{\textbf{Loc.err.} \\ (50 cm) $\downarrow$} &
\textbf{Avg.err.} $\downarrow$ &
\textbf{Pose Dist} $\downarrow$ \\
\midrule
w/o Perturbation  & 0.3006 & 0.1982 & 0.3491  & 0.0618 \\
w/o ControlNet  & 0.1383 & 0.0336 & 0.0754  & 0.1622 \\
\textbf{Ours} & \textbf{0.0246} & \textbf{0.0076} & \textbf{0.0250} & \textbf{0.0210} \\
\bottomrule
\end{tabular*}

\label{tab:ablation_humanml3d}
\end{table}

\begin{figure}[htbp]
    \centering
    \includegraphics[width=1\linewidth]{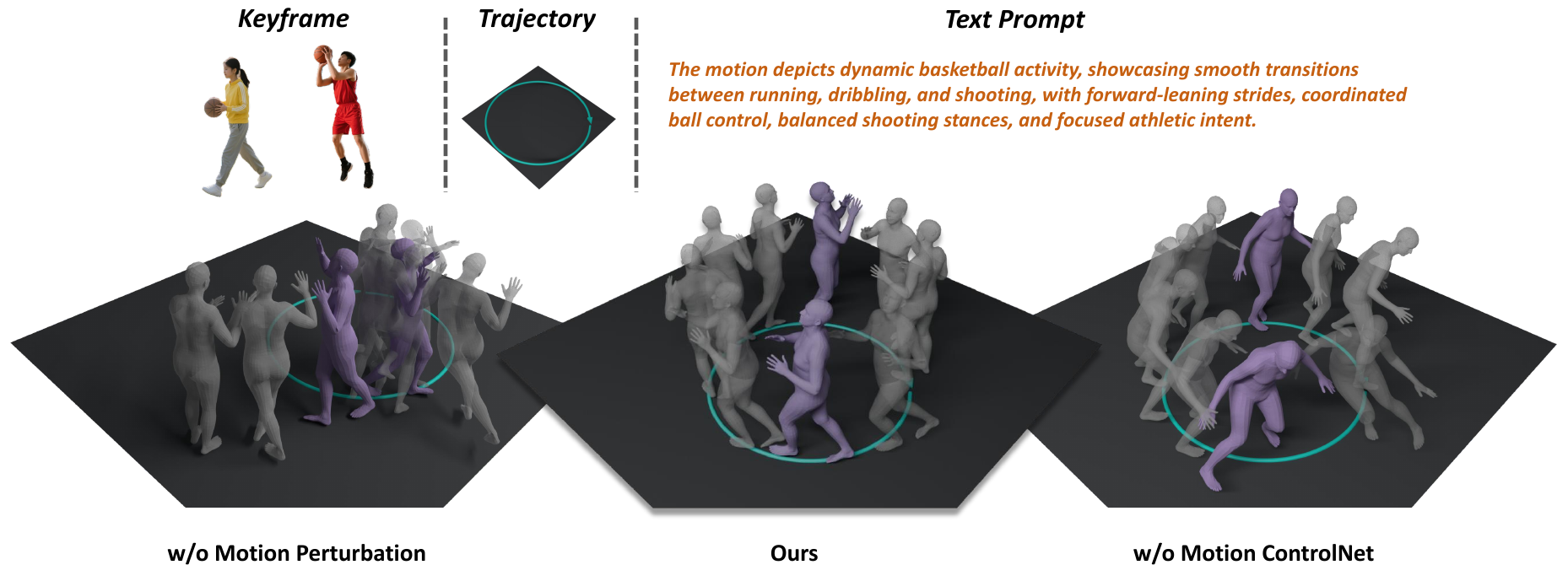}
    \caption{\textbf{Ablation results. }}
    \label{fig:ablation}
\end{figure}

\textbf{Decoupled Strategy.} As shown in Table~\ref{tab:comparison_humanml3d} and Table~\ref{tab:posedist_humanml3d}, the decoupled design improves motion quality and control accuracy across nearly all metrics, demonstrating its overall effectiveness in enhancing model performance.

\textbf{Motion Perturbation.} Table~\ref{tab:ablation_humanml3d} shows that enabling Motion Perturbation yields a $>$90\% reduction in trajectory errors (\textit{Traj.err.}, \textit{Loc.err.}, and \textit{Avg.err.}), underscoring its necessity. Visual results in Figure~\ref{fig:ablation} (left) further corroborate that omitting this module causes severe trajectory deviation.

\textbf{Perturbation + ControlNet.} Table~\ref{tab:ablation_humanml3d} shows that \textit{Pose Dist} degrades by 2.94$\times$ without Perturbation and 7.72$\times$ without ControlNet, highlighting the latter's criticality. Figure~\ref{fig:ablation} (right) visually corroborates these misalignments.

\textbf{Rotation-Augmented Training.} Table~\ref{tab:ablation_rotation} shows that applying random rotations during training improves robustness to rotational variations at inference.

\begin{table}[htbp]
\caption{\textbf{Ablation studies on rotation.} \textit{Eval Rotations} indicates whether random rotations are applied during evaluation.}
\centering

\small
\setlength{\tabcolsep}{1.5pt}
\begin{tabular*}{\linewidth}{@{\extracolsep{\fill}} llccc}
\toprule
\textbf{Method} &
\makecell{\textbf{Eval} \\ \textbf{Rotations}} &
\makecell{\textbf{FID} \\ $\downarrow$} &
\makecell{\textbf{R-precision} \\ (Top-3) $\uparrow$} &
\makecell{\textbf{Diversity} \\ $\rightarrow$}  \\
\midrule
Ours(w/ rotations) & w/o Rotations & 0.239 & \textbf{0.678} & 9.686 \\
Ours(w/o rotations) & w/o Rotations & \textbf{0.177} & 0.672 & \textbf{9.616} \\
\midrule
Ours(w/ rotations) & w/ Rotations & \textbf{0.238} & \textbf{0.671} & \textbf{9.642}\\
Ours(w/o rotations) & w/ Rotations & 0.594 & 0.635 & 9.032 \\
\bottomrule
\end{tabular*}

\label{tab:ablation_rotation}
\end{table}

\section{Conclusion}
In this work, we identify that while unified conditioning mechanisms excel in professional workflows, they struggle with heterogeneous geometric constraints arising from unaligned, image-derived inputs. To resolve the inherent conflict between absolute trajectories and relative poses, we propose \textbf{TPMDM}, which employs a decoupled parallel strategy for explicit coordinate alignment. Furthermore, we introduce \textbf{IKMo}, a holistic framework synergizing TPMDM with VLM agents to bridge the semantic-geometric gap. Experiments on HumanML3D and KIT-ML demonstrate that our method establishes a new state-of-the-art in the challenging scenario of joint control with heterogeneous trajectories and keyframe poses.

\bibliographystyle{IEEEtran}
\bibliography{icme2026references}

\end{document}